\documentclass{jfm}
\usepackage{graphicx}
\usepackage{subcaption}
\usepackage{color}
\usepackage{tikz}

\newcommand{\myline}[1]{\raisebox{2pt}{\tikz{\draw[-,{#1},solid,line width = 1.5pt](0,0) -- (4mm,0);}}}

\definecolor{undef_c}{rgb}{0.0504,0.0298,0.5280}
\definecolor{steady_def_c}{rgb}{0.3656,0.0030,0.6499}
\definecolor{flap_def_c}{rgb}{0.6107,0.0902,0.6200}
\definecolor{large_c}{rgb}{0.7964,0.2780,0.4713}
\definecolor{chaotic_c}{rgb}{0.9283,0.4730,0.3261}
\definecolor{def_mode_c}{rgb}{ 0.9936,0.7018,0.1845}

\shorttitle{}
\shortauthor{A. Goza}

\title[Flapping of a nonuniform-stiffness inverted flag]{Flow-induced flapping of an inverted flag with non-uniform stiffness distribution}

\author{Andres Goza\aff{1}
  \corresp{\email{agoza@illinois.edu}} }

\affiliation{\aff{1}Department of Aerospace Engineering, University of Illinois at Urbana-Champaign, IL, USA}

\begin{document}

\maketitle

\begin{abstract}
We perform high-fidelity, two-dimensional (2D), fluid-structure interaction (FSI) simulations at a Reynolds number of $Re=200$ of uniform flow past an inverted flag (i.e., clamped at its trailing edge). The inverted flag system can exhibit large-amplitude flapping motions (on the order of the flag length) that can be converted to electricity via, e.g., piezoelectric materials. We investigate the effect of structural nonuniformity in altering the FSI dynamics compared with the uniform-stiffness scenario that has been thoroughly characterized. We consider linear, quadratic, and cubic stiffness distributions, and demonstrate that the FSI dynamics mirror those of a uniform-stiffness flag with an appropriately defined effective stiffness. We show that this effective stiffness can be computed simply via analysis of an \emph{in-vacuo} Euler-Bernoulli beam. When expressed in terms of the effective stiffness, the FSI dynamics of the nonuniform-stiffness flag exhibit the same regimes---with many similarities in the detailed dynamics---as a uniform-stiffness flag. This study opens questions about (i) what the optimal stiffness distribution is for, e.g., energy harvesting capacity, and (ii) how to use nonuniform (and possibly time-varying) stiffness distributions to control the flag dynamics towards a desired state. 
\end{abstract}

\section{Introduction}

Uniform flow past a deformable flag is commonly studied as an archetypal fluid-structure interaction (FSI) system for a range of phenomena that are pervasive to natural and engineered systems. Studies of the inverted-flag system, where the flag is clamped at its trailing edge with respect to the oncoming flow, have revealed that the inverted-flag system exhibits rich dynamics with multiple behavioral regimes. One of these regimes involves large-amplitude flapping (on the order of the flag's length), which is associated with large bending strains that provide potential for renewable energy harvesting by, \emph{e.g.}, coupling the flag to a piezoelectric material \citep{Shoele2016,orrego2017harvesting}. 

The parameters that govern the dynamics of the flag system are the dimensionless mass and stiffness as well as the Reynolds number, defined here (respectively) as 
\begin{equation}
    M = \frac{\rho_s h}{\rho_f L}, \; K = \frac{D}{\rho_f U^2 L^3}, \; Re = \frac{\rho_f U L}{\mu},
    \label{eqn:param_def}
\end{equation}
where $\rho_f$ ($\rho_s$) is the fluid (structure) density, $U$ is the freestream velocity, $L$ is the flag length, $\mu$ is the shear viscosity of the fluid, $h$ is the flag thickness, and $D$ is the flexural rigidity of the flag. A variety of behavioral regimes can be observed by varying these dimensionless parameters (with the dimensionless stiffness $K$ playing the most significant role in determining the dynamics). 

For large stiffness values, an undeformed equilibrium state with the flag remaining undeflected has been found experimentally, computationally, and theoretically. This equilibrium has been shown to become unstable with decreasing stiffness due to a divergence instability \citep{Kim2013,Gurugubelli2015,Sader2016a}. This instability gives way to a small-deflection stable state for low Reynolds numbers, $O(100-1{,}000)$ \citep{Ryu2015,Gurugubelli2015}. This deformed stable state was shown to be a formal equilibrium of the full FSI system, and was found to give way with a further decrease in stiffness to small-deflection flapping caused by a supercritical Hopf bifurcation \citep{Goza2018}. These small-deflection regimes have not been reported in higher Reynolds numbers experiments, though this may be due in part to the narrow range of stiffness values over which this regime is observed. 

With further decrease in stiffness, large-amplitude flapping occurs with flapping amplitudes that are commensurate with the flag length. This regime has been linked for low mass ratios $M<1$ to a vortex-induced vibration (VIV; see, \emph{e.g.}, \citet{Sarpkaya2004,Williamson2004}), with flapping initiated by leading-edge separation \citep{Gurugubelli2015} and subsequent synchronization occurring between the integrated fluid forces and flapping dynamics occurring near a Strouhal number of $0.2$ \citep{Sader2016a}. Disruptions to this VIV behavior have been observed. Experimental and numerical investigations have identified that even when serrations and splitter plates were used to disrupt key flow structures, flapping persisted at different frequencies and produced distinct vortex shedding processes \citep{tavallaeinejad2020flapping,gurugubelli2019large}. Moreover, for large mass ratios ($M>1$) flapping is qualitatively distinct from the classical VIV phenomena, with simulations highlighting different flapping-vortex shedding interactions \citep{Goza2018} and theoretical work demonstrating that a quasi-steady fluid model---devoid of vortex shedding---predicted many features of the heavy-flag flapping dynamics \citep{tavallaeinejad2020instability}.

Further decrease in stiffness yields to a de-synchronization between flapping and vortex shedding \citep{Goza2018} and subsequent chaotic flapping, which has been found experimentally at high Reynolds numbers of $O(10^4)$ \citep{Sader2016a} and computationally at the lower Reynolds number of $200$ \citep{Goza2018}. \citet{Goza2018} characterized this regime as one with an attractor involving (seemingly) random switching between large-amplitude flapping and the deformed mode. The authors demonstrated that chaotic flapping does not occur at low Reynolds numbers $\lesssim 50$ (where classical bluff-body vortex shedding is absent) or for massive flags ($M>1$), which suggests that chaos arises from a de-synchronization/breakdown of VIV behavior in large-amplitude flapping of light flags, but before the deformed mode becomes a global attractor.

A further decrease in stiffness results in the deformed mode regime, in which there is small-amplitude flapping occurs about a large mean-deflection state. This regime has been observed experimentally \citep{Kim2013} and computationally  \citep{Ryu2015,Gurugubelli2015,Shoele2016}. For Reynolds numbers below $\lesssim50$ (where classical bluff-body vortex shedding ceases), the regime is an equilibrium state of the FSI system \citep{Goza2018}. At higher Reynolds numbers, flapping occurs near the classical bluff-body frequency scaling of 0.2 \citep{Shoele2016}.

The sensitivity of the large-amplitude flapping regime highlights a challenge for energy harvesting: small changes in dimensionless stiffness ($K$, achievable through, \emph{e.g.}, a change in the oncoming flow velocity) can elicit a transition from large-amplitude flapping to a different regime that is less useful for energy capture. A possibility for addressing this sensitivity is to manipulate the flag's structural properties (made possible by advances in modern materials) to alter the flag dynamics. To inform how to tune these material properties, a key question is what role structural nonuniformity plays in dictating the dynamics of the inverted-flag system. We will answer this question here in the case of a flag with a linear, quadratic, and cubic stiffness distribution. Our studies use high-fidelity two-dimensional (2D) simulations at $Re=200$, $M=0.5$. We demonstrate that the dynamics of the nonuniform-stiffness flag can be predicted by a uniform-stiffness flag, provided that an effective stiffness is utilized. We further show that this effective stiffness is different from the mean stiffness, and can be predicted by a simple calculation involving an Euler-Bernoulli beam in a vacuum. When expressed in terms of the effective stiffness, the FSI dynamics of the nonuniform-stiffness flag are shown to exhibit the same regimes---with many similarities in the detailed dynamics---as a uniform-stiffness flag. 

We restrict our attention to the aforementioned mass ratio $M$ and Reynolds number $Re$. The relative robustness of the dynamics to these two parameters suggests that features of our conclusions will hold for other flag masses and Reynolds numbers. We also do not investigate 3D effects here. The commonality between 3D experiments and 2D simulations highlighted above as well as systematic studies of aspect ratio suggest that the majority of 2D phenomena persist in 3D, provided that the aspect ratio is not too small \citep{Sader2016b,tavallaeinejad2021dynamics}. Similarly, the inclination angle of the flow with respect to the flag is not considered here, but can alter the flag dynamics considerably \citep{Shoele2016,huertas2021dynamics}. An investigation into the role of 3D phenomena and the initial inclination angle is left to future study.

\section{Problem statement and numerical method}
\label{sec:methods}

In this work, we consider an inverted-flag system in which the flag is clamped at its leading edge with respect to the oncoming flow. The dimensionless mass and Reynolds number are fixed in this study at $M=0.5$ and $Re=200$, respectively. Drawing inspiration from \citet{floryan2020distributed}, who studied the role of structural nonuniformity in bio-inspired locomotion, we express the dimensionless stiffness distribution in terms of Legendre polynomials as $K(s) = 2\overline{K} + \sum_{j=1}^n c_j L_j(s)$, where $s\in[-1,1]$ is a parametrization of the flag (non-dimensionalized by the flag length $L$), $\overline{K}$ is the mean stiffness, and $L_j$ is the $j^{th}$ Legendre polynomial. 
This stiffness distribution uses the same non-dimensionalization as in (\ref{eqn:param_def}). In this work, we consider linear, quadratic, and cubic stiffness distributions, corresponding to $n=1,$ $2$, and $3$, respectively. The benefit of the representation via Legendre polynomials is that  $\langle L_j, 1\rangle = 0$, where $\langle\cdot, \cdot\rangle$ is the standard (unweighted) inner product on the interval $[-1,1]$. That is, varying the coefficients $c_1$, $c_2$, and $c_3$ does not affect the mean stiffness $\overline{K}$. This approach allows for the effect of stiffness distribution to be investigated without inheriting effects from changing the mean stiffness. The only restriction on $c_1$, $c_2$, and $c_3$ is thus that $K(s)>0$  $\forall s\in[-1,1]$.

\begin{figure}
    \centering
    \includegraphics[width=0.9\textwidth]{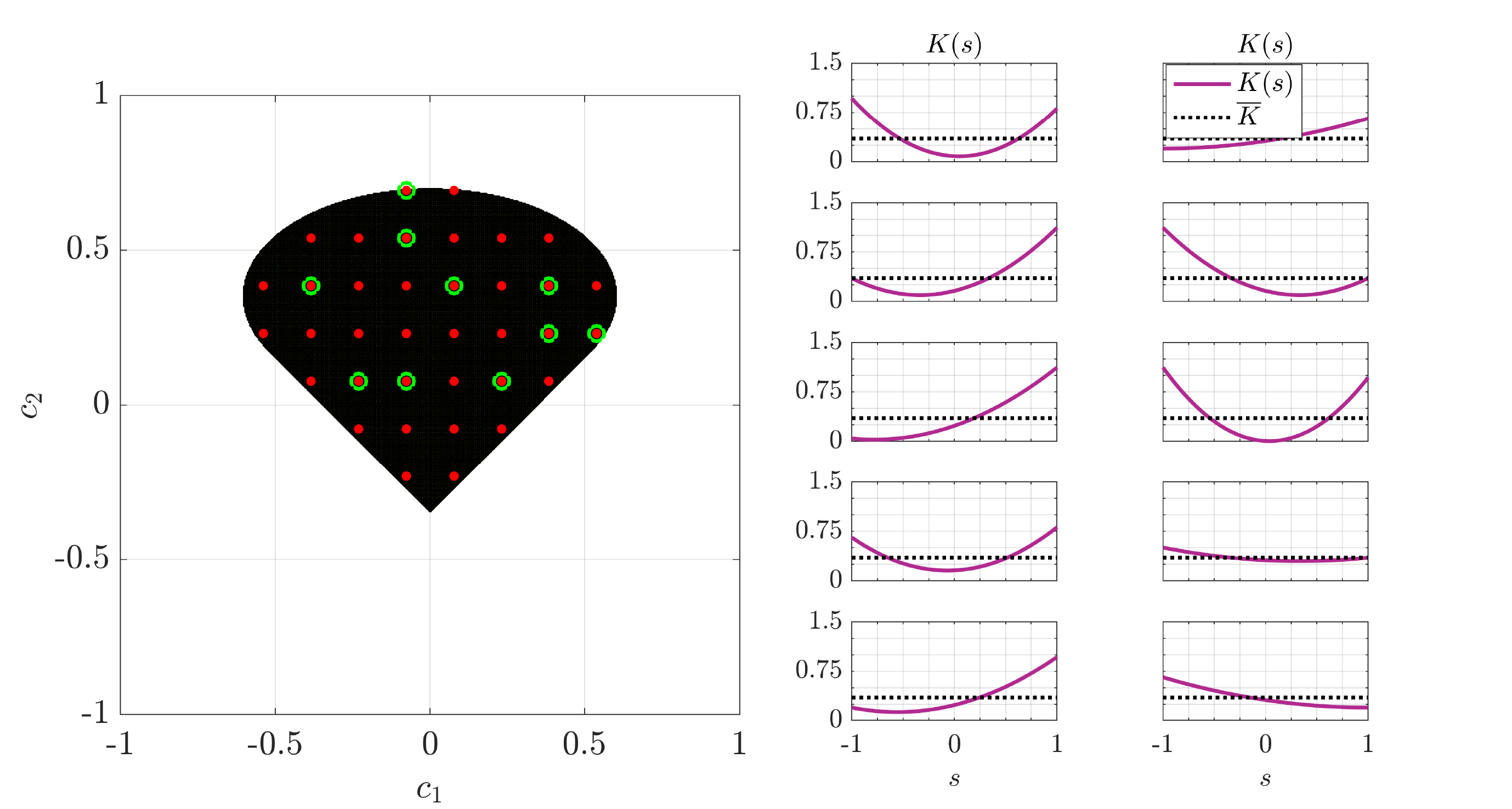}
    \caption{Example for $\overline{K}=0.35$, $n=2$ of how the coefficients $c_1$, $c_2$, and $c_3$ were selected to define the nonuniform stiffness distribution $K(s)$. Left subplot: feasible set of coefficients (shaded region) and coefficients that were used for the simulations (red markers). Right subplot: stiffness distributions for a subset of the utilized coefficients (green open circles in the left subplot). The coefficients used for other values of mean stiffness and polynomial order were obtained through the same process. }
    \label{fig:sample_dist}
\end{figure}

To meaningfully sample the parameter space associated with the various coefficients, we focus on mean stiffness values of $\overline{K}=0.15$, $0.25$, $0.35$, and $0.45$. These mean stiffnesses correspond in the uniform-distribution studies to the deformed mode, chaotic, large-amplitude, and small-deflection flapping regimes, respectively, for this choice of mass ratio and Reynolds number. For each mean stiffness $\overline{K}$ and polynomial degree $n$, the coefficients $c_1$, $c_2$, and $c_3$ are varied as follows. We first construct the feasible set of coefficients for the given mean stiffness $\overline{K}$ and polynomial degree $n$. We then perform simulations using a uniformly distributed set of coefficients within the feasible set. Figure \ref{fig:sample_dist} shows the feasible set and values of $c_1$, $c_2$ considered for $\overline{K}=0.35$ and $n=2$. The specific number of simulations varied between fifteen and forty for each choice of $\overline{K}$ and $n$ based on the size of the feasible set. In total, seventy four, eighty three, and eighty nine runs were performed for $n=1$, $2$, and $3$, respectively.

All simulations are run using the immersed boundary method of \citet{Goza2017}. The method has been validated on a number of flapping flag problems for flags in both the conventional and inverted configuration. The method was also used to identify physical mechanisms in the inverted-flag system for a flag of uniform stiffness \citep{Goza2018}. The simulation parameters for the present study are the same as in the cited reference; a convergence test therein demonstrates the suitability of those simulation parameters (a separate convergence test was run using the parameters for the present study and found to give similar results---this outcome is expected, as the dynamics for this non-uniform stiffness setting will be shown below to be commensurate with those from the uniform-stiffness case). 

For all simulations, the flag is initialized in the undeformed configuration, and dynamics are triggered by introducing a small body force at early time. To ensure that long-term behavior is captured, the simulations are run for a minimum of 150 convective time units (for the cases exhibiting chaotic dynamics, simulations are run for 300 convective time units). In the cases exhibiting limit-cycle dynamics, this final time amounts to several dozen periods being captured; for reference, in the left two columns of figure \ref{fig:uni_summary} below, the $x$-limits are cut off at 100 convective time units. 

\section{Summary for a uniform flag and intuition behind effective stiffness}

\begin{figure}
    \centering
    \hspace*{-11.5mm}
    \includegraphics[width=1.18\textwidth]{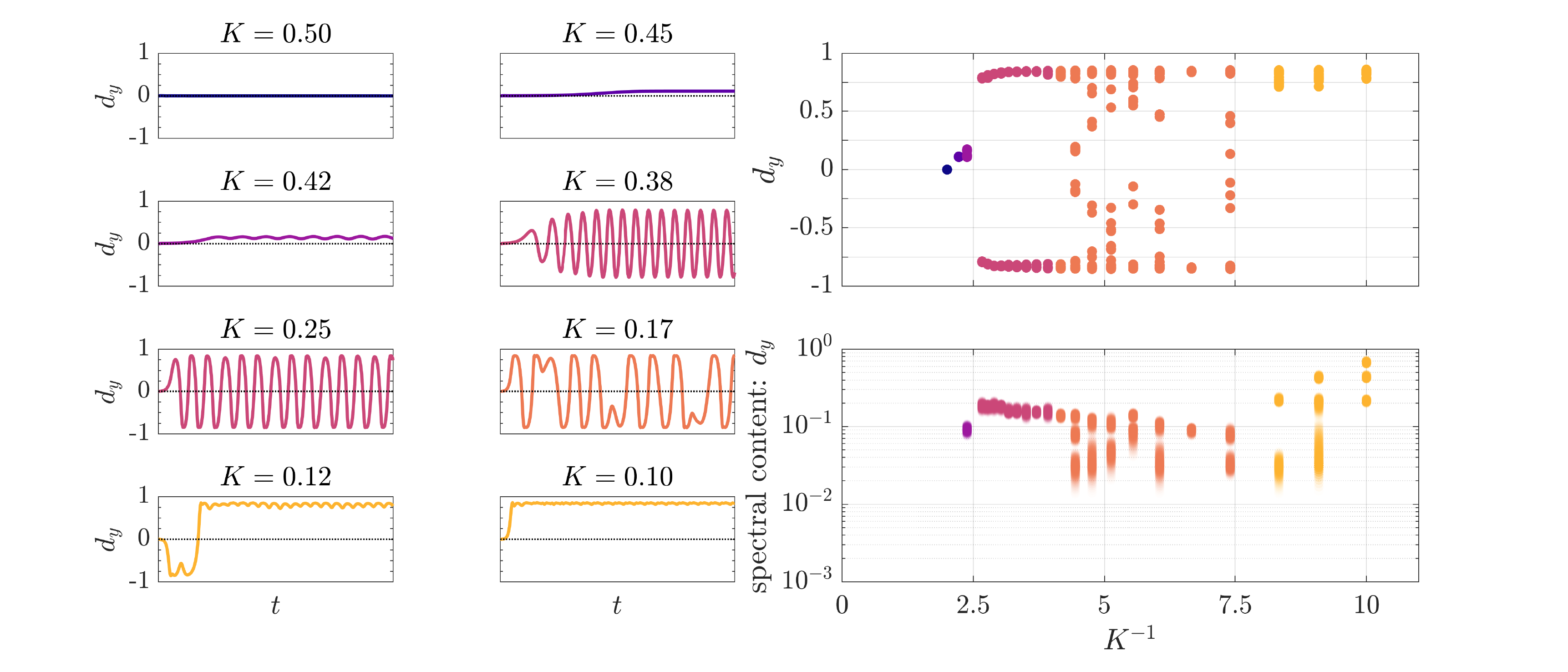}
    \caption{Left two columns: transverse displacement of the flag's leading edge, $d_y$, as a function of time for the undeformed equilibrium (\protect\myline{undef_c}), small-deflection equilibrium (\protect\myline{steady_def_c}), small-deflection flapping (\protect\myline{flap_def_c}), large-amplitude flapping (\protect\myline{large_c}), chaotic flapping (\protect\myline{chaotic_c}), and the deformed mode (\protect\myline{def_mode_c}) regimes. Regime transitions were triggered by decreasing the dimensionless stiffness $K$. Third column: Alternative view of the regimes shown as a bifurcation diagram (top row) and frequency content of the tip displacement versus stiffness $K$ (bottom row). For the bifurcation diagrams, at a given stiffness the markers show the leading-edge displacement at zero velocity. (For example, for a stiffness in the chaotic regime, the markers at various $d_y$ values reflect irregular direction changes; \emph{c.f.}, the corresponding timetrace plot in the second column). The frequency information in the bottom right subplot provides, for each $K$, the frequency peaks from a power spectral density analysis with at least 20\% of the frequency peak with maximum energy. To indicate the spread of energy about each frequency peak, the color extends from the peak across neighboring frequencies with at least 5\% of the peak energy. (For example, for $K^{-1}\approx5$ there are two dominant frequencies, near $0.03$ and $0.1$, and there is considerable spread in energy across neighboring frequencies for the peak near $0.03$).} 
    \label{fig:uni_summary}
    \vspace*{-0.0cm}
\end{figure}

To build context for the ensuing non-uniform stiffness results, we review here the salient dynamics of a uniform-stiffness inverted flag. Figure \ref{fig:uni_summary} demonstrates the various regimes observed by decreasing the stiffness $K$. The frequency content (bottom right subplot) highlights that the transition to large-amplitude flapping coincides (at this mass ratio of $M=0.5$) with a jump from low flapping frequency to one that is synchronized to vortex shedding at the classical bluff body scaling of 0.2. This is one of the markers of the VIV behavior that the system exhibits \citep{Sader2016a}. With decreasing stiffness, the flag frequency decreases steadily, signaling a de-synchronization between flapping and vortex shedding. Eventually, the de-synchronization disrupts large-amplitude flapping and triggers chaotic flapping with broadband frequency content. Finally, with further decreases in stiffness, the flag enters into the deformed mode regime, and the dominant frequency is near 0.2 and its harmonics (though there are relatively small peaks for some stiffnesses at lower frequencies, $\lesssim0.04$). The prominent frequency signatures of 0.2 reflect that the small-amplitude flapping is driven by forcing from canonical bluff-body vortex shedding \citep{Shoele2016}. See \citet{Goza2018} for a more detailed characterization of the uniform-stiffness dynamics for this choice of Reynolds number, mass ratio, and collection of stiffness values.

\begin{figure}
    \centering
    \hspace*{-11.5mm}
    \includegraphics[width=1.18\textwidth]{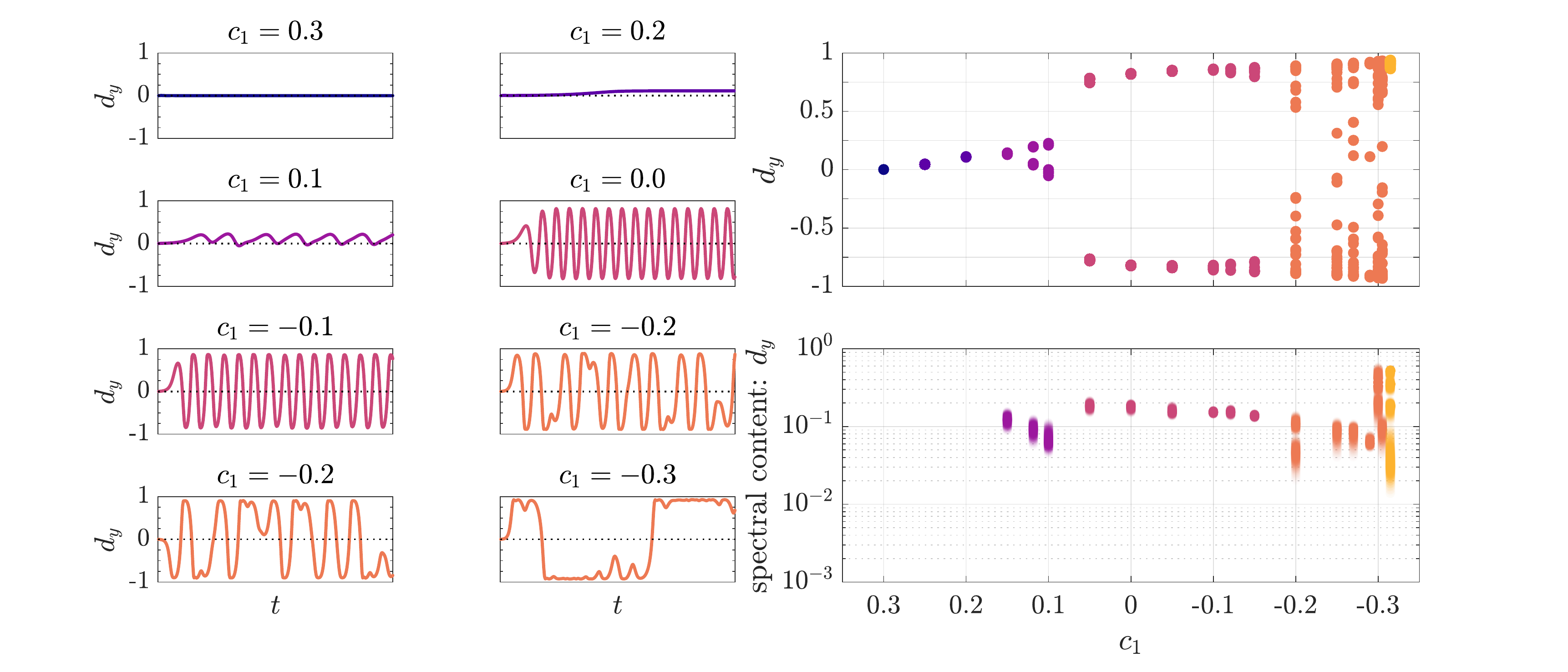}
    \caption{Analog of figure \ref{fig:uni_summary} except $c_1$ is varied, not $K$.}
    \label{fig:gen_results_fig}
    \vspace*{-0cm}
\end{figure}

We now provide archetypal results in figure \ref{fig:gen_results_fig} for a linear stiffness distribution ($c_2=c_3=0$) with mean stiffness $\overline{K}=0.35$ and various $c_1$ values. Note that since $L_1(x)=x$, $c_1$ is the slope, with negative (positive) values corresponding to stiffness being more isolated towards the leading (trailing) edge. The figure demonstrates that $c_1$ acts as a bifurcation parameter, analogous to the role that $K$ plays for the uniform stiffness case; \emph{c.f.}, figure \ref{fig:uni_summary}. Indeed, the linearly distributed case of figure \ref{fig:gen_results_fig} shows that the same regimes (undeformed, small-deflection equilibrium, small-deflection flapping, large-amplitude flapping, and deformed mode) exist, and that the bifurcation sequence is the same with decreasing $c_1$ as for decreasing $K$. Moreover, the frequency content of the linearly distributed case (bottom right subplot of figure \ref{fig:gen_results_fig}) demonstrates that the same dynamics are present as in the uniform-stiffness case: small-amplitude flapping of low frequency gives way to VIV flapping at large amplitudes, and a subsequent de-synchronization of flapping and vortex shedding yields chaotic flapping and eventually the deformed-mode regime characterized by some broadband low frequency dynamics, but even stronger frequency peaks near $0.2$ and its integer harmonics. 
These facts suggest that the flag with a linear stiffness distribution acts as a uniform stiffness flag with an effective stiffness different from the mean stiffness: for a given mean stiffness, flags with a stiffer trailing edge ($c_1>0$) act effectively stiffer and flags with a more flexible trailing edge ($c_1<0$) act more flexible than a uniform stiffness flag with stiffness $K\equiv\overline{K}$. 

\begin{figure}
    \centering
    \includegraphics[width=1\textwidth]{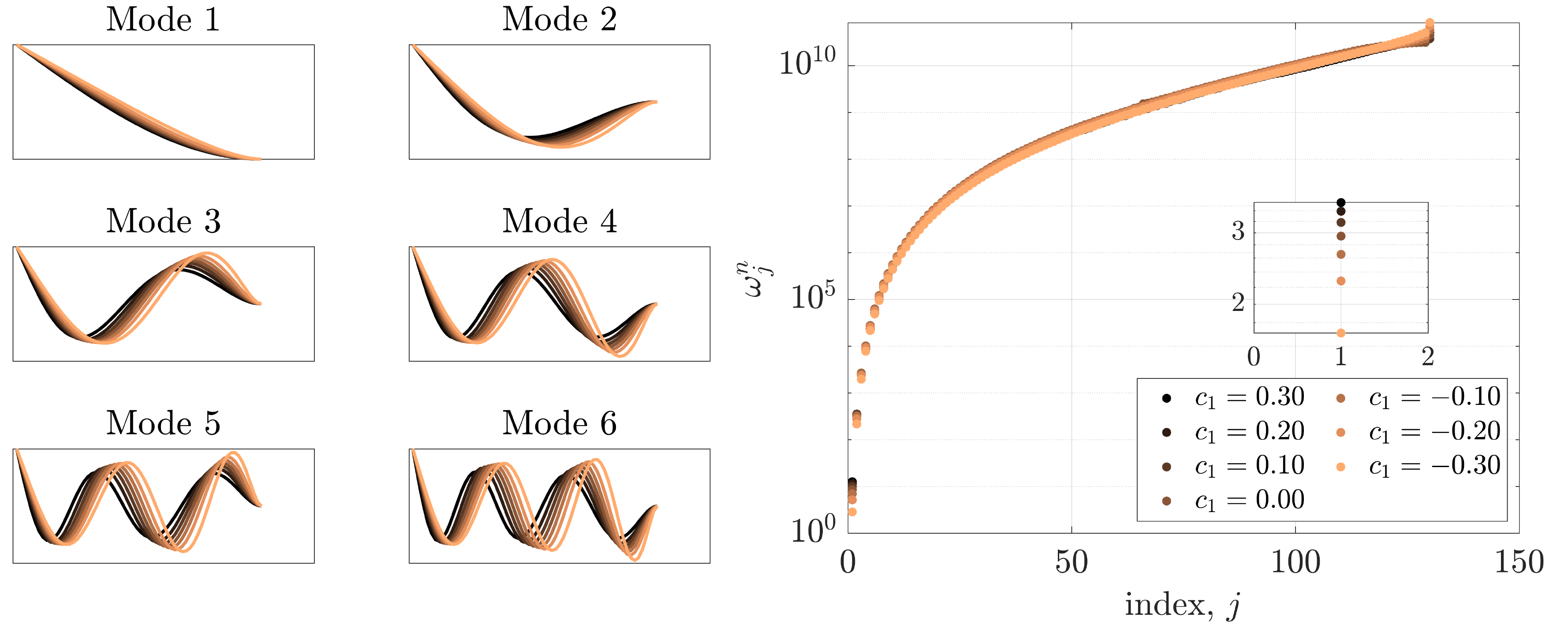}
    \caption{Leading eigenvectors (left two columns) and natural frequencies, $\omega^n_j$, versus index $j$ (right column) for an \emph{in-vacuo} Euler-Bernoulli beam with mean stiffness $\overline{K}=0.35$ and linear stiffness distribution ($c_2=c_3=0$). The insert in the natural frequency plot shows the natural frequency of the leading mode, index $j=1$.}
    \label{fig:EB_vac}
\end{figure}

We will define this effective stiffness from an analogous uniform flag. To build intuition, we first consider an Euler-Bernoulli beam in a vacuum with mean stiffness $\overline{K}=0.35$ and linear stiffness distribution ($c_2=c_3=0$). This beam model is a linearized variant of the geometrically nonlinear structural model used in our numerical method, described in section \ref{sec:methods}. We show in figure \ref{fig:EB_vac} the leading eigenvectors and eigenvalues (computed by an eigendecomposition of a prototypical finite element discretization) for various values of $c_1$. The figure demonstrates that while the second through sixth eigenvectors vary in shape with varying $c_1$, to reasonable approximation the leading eigenvector does not. The similarity of the leading eigenvector with varying $c_1$ suggests that the dynamics can be represented by a uniform-stiffness structure, as the FSI inverted-flag system is dominated by first mode dynamics. The right subplot shows that the natural frequency increases (i.e., the effective stiffness increases) as stiffness is focused more towards the clamped (trailing) edge. This trend suggests a procedure for computing the effective stiffness, which we employ here. Given a stiffness distribution, we define the effective stiffness as the stiffness for a uniform-property flag that has the same leading natural frequency as the non-uniform stiffness flag of interest. To indicate why focusing stiffness towards the trailing edge (leading edge) leads to an effectively stiffer (more flexible) flag, we note that the strain energy within the flag is given by $\int_{-1}^1 K(s) \kappa(s) ds$, where $\kappa(s)$ represents the curvature of the flag. The material stiffness $K(s)$ thus contributes more to the strain energy---and thereby overall stiffness---in areas where curvature is more pronounced. The clamped end is a natural location to expect these regions of high curvature.

We note that the interaction of the flag with the surrounding flow will alter the natural frequency of the FSI system. We do not argue here that the vacuum-scaled natural frequency is the correct natural frequency for the FSI system, but instead that the effect the flow has on the non-uniform flag is similar to its effect on a uniform flag when an appropriately defined effective stiffness (distinct from the mean stiffness) is used.

\section{Representing dynamics with the effective stiffness}
\label{sec:effective_stiffness}

\begin{figure}
    \centering
    \includegraphics[width=1\textwidth]{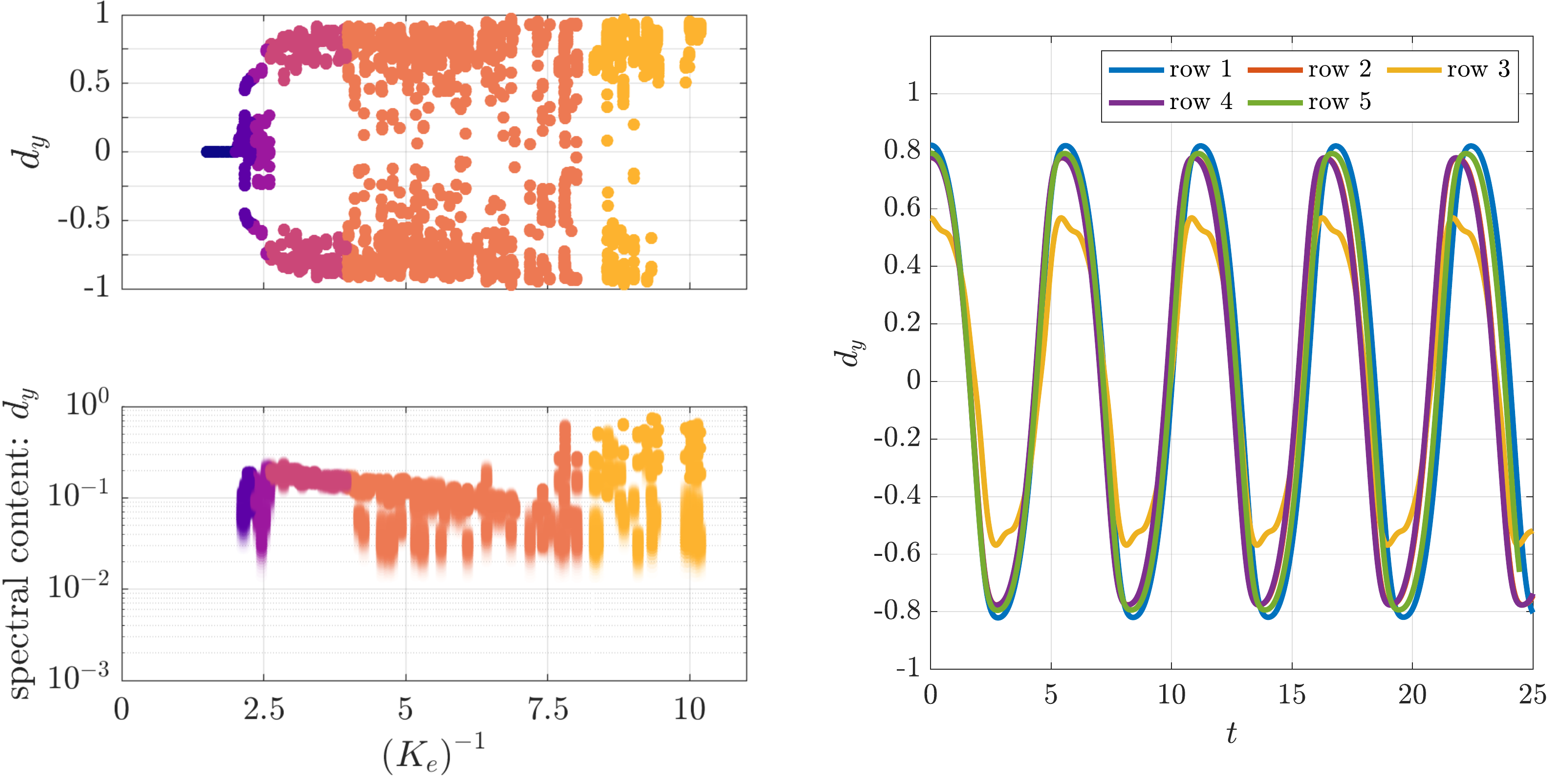}
    \caption{Left column: analog of the right column of figure \ref{fig:uni_summary} for a variety of linear, quadratic, and cubic non-uniform stiffness distributions. For ease of comparison with the uniform-stiffness case, the color scheme is the same as in figure \ref{fig:uni_summary}. Right column: time trace of $d_y$ for five cases with an effective stiffness of $K_e\approx0.35$ (the legend provides the corresponding row in figure \ref{fig:K_p35_snaps}, where parameter details are given). }
    \label{fig:bif_all_and_timetrace}
\end{figure}

We now utilize the definition of the effective stiffness, $K_e$, to represent the dynamics for linear, quadratic, and cubic stiffness distributions in figure \ref{fig:bif_all_and_timetrace}. The left column of figure \ref{fig:bif_all_and_timetrace} is created from two hundred forty six runs, using all three distributions and mean stiffness values of $\overline{K}=0.15$, $0.25$, $0.35$, and $0.45$ (corresponding in the uniform case to the deformed mode, chaotic, large-amplitude, and small-deflection flapping regimes, respectively). The left column of the figure demonstrates that the regime transitions occur at nearly the same $K_e$ as for the uniform-stiffness case (c.f., figure \ref{fig:uni_summary}). In addition, the frequency content suggests that the transitions transpire by the same mechanisms with decreasing effective stiffness---a jump in frequency to large-amplitude VIV flapping, de-synchronization that triggers chaos and broadband frequency dynamics, and the eventual settling of the flag to one side in a deformed-mode state characterized by small-amplitude flapping due to flow forcing from bluff-body vortex shedding.

The most notable exceptions to the predictive capabilities of $K_e$ are the occurrence of chaotic dynamics where the deformed mode is predicted, and the difference of flapping amplitude across the large-amplitude flapping regime. For the former exception, the corresponding frequency content demonstrates that there are prominent signatures at 0.2 and its harmonics, which reflects the fact that the flag spends significant time in a quasi-deformed mode state before flapping to the other side of the centerline. 

\begin{figure}
    \centering
    \includegraphics[width=0.9\textwidth]{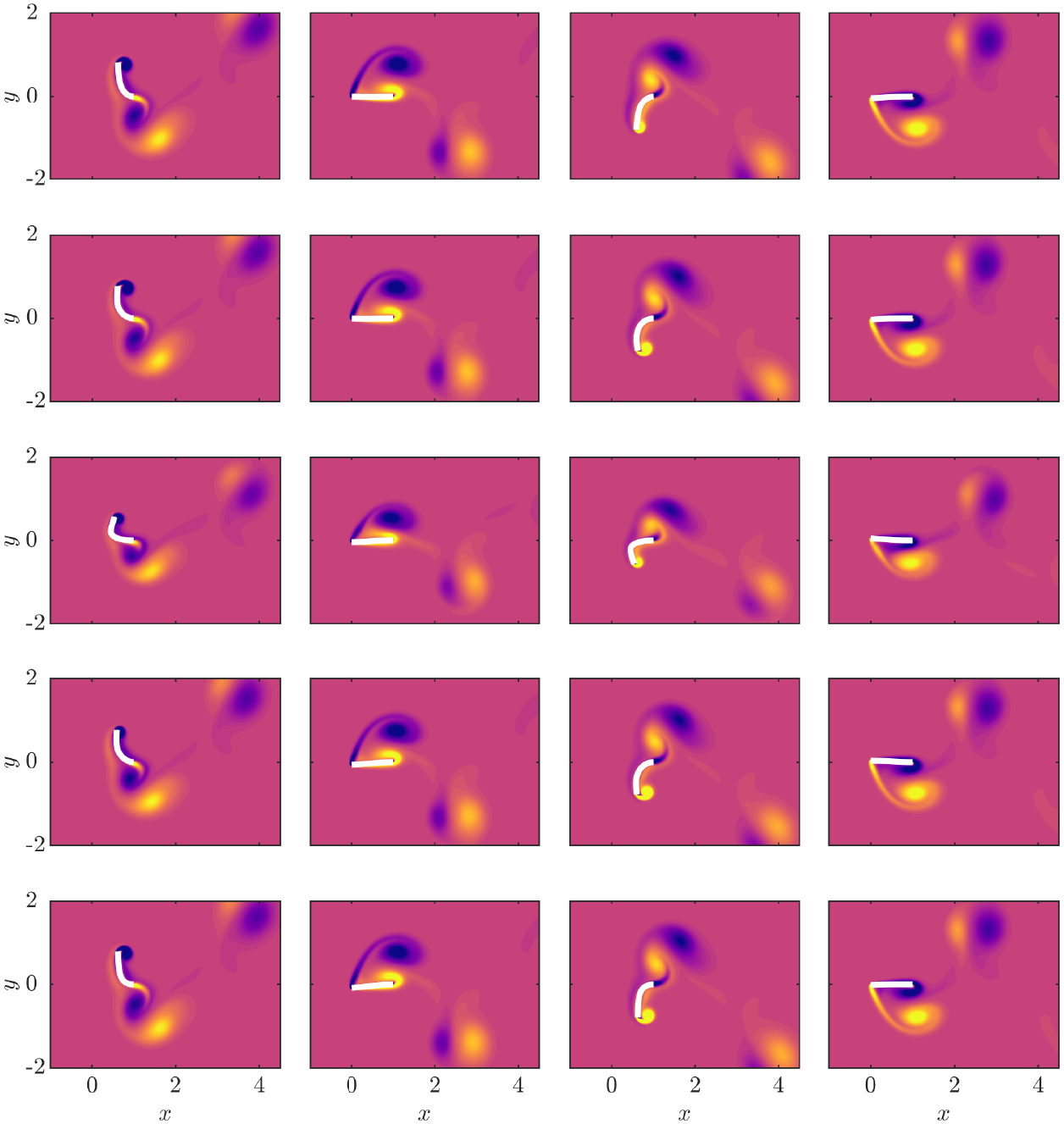}
    \caption{Each row provides snapshots of vorticity, $\omega$, at four equally spaced time instances across a flapping period for a flag with effective stiffness near $K_e=0.35$. First row: $\overline{K}=0.35$, $c_1=c_2=c_3=0$; second row: $\overline{K}=0.25$, $c_1=0.1765,$ $c_2=0.0588$, $c_3=0$, $K_e=0.350$; third row: $\overline{K}=0.45$, $c_1=0.5172,$ $c_2=0.7241$, $c_3=0$, $K_e=0.351$; fourth row: $\overline{K}=0.45$, $c_1=0.1765,$ $c_2=0.0588$, $c_3=-0.05882$, $K_e=0.351$; fifth row: $\overline{K}=0.45$, $c_1=-0.1579,$ $c_2=0.1579$, $c_3=0.2632$, $K_e=0.351$. There are thirty four vorticity contour levels over $\omega\in[-10,10].$ }
    \label{fig:K_p35_snaps}
\end{figure}

Within the large-amplitude flapping regime, for all but one of the cases the dynamics evolve onto a limit cycle with a fixed amplitude (to within $5\%$). Thus, the variation in tip displacement $d_y$ is because different stiffness distributions can lead to different flapping amplitudes---a notable contrast from the nearly constant amplitude across this regime for the uniform case. The periodic dynamics and variation in flapping amplitude are demonstrated further in the right column of figure \ref{fig:bif_all_and_timetrace}. Yet, despite this apparent difference from the uniform case, there remain marked similarities in the detailed dynamics. Figure \ref{fig:K_p35_snaps} provides vorticity snapshots at equally spaced instances during a flapping cycle for the five stiffness distributions from the timetrace plot in the right column of figure \ref{fig:bif_all_and_timetrace}. All five cases produce similar vortex dynamics, with a leading-edge vortex (LEV) and trailing-edge vortex (TEV) formed when the flag is at its peak and minimal displacement values, respectively. Moreover, a P+P wake structure forms and advects in nearly identical fashion. At the same time, for the instances of peak displacement the case with smaller flapping amplitude leads to significant additional bending near the chordwise center. This diversity of bending within common FSI dynamics raises questions about what bending distributions are most useful for energy capture within large-amplitude flapping.

\section{Conclusions}

We performed high-fidelity 2D FSI simulations of flow past an inverted flag with linear, quadratic, and cubic distributions. The stiffness distribution was defined using Legendre polynomials so that stiffness distribution could be systematically varied without affecting the mean stiffness. We demonstrated that the dynamics of the non-uniform flag could be represented as those of a uniform flag with an appropriately defined effective stiffness. This effective stiffness was computed in a straightforward manner using an \emph{in-vacuo} Euler-Bernoulli beam: the effective stiffness was defined as the stiffness of a uniform flag with the same leading natural frequency as that of the non-uniform stiffness flag. When cast in terms of this effective stiffness, the same regimes (with minor exceptions) occurred for linear, quadratic, and cubic distributions as in the uniform case. Moreover, many of the same dynamical details persisted. Large-amplitude flapping continued to be characterized by synchronization between vortex shedding and flapping, chaotic flapping was triggered by a de-synchronization of these phenomena, and the deformed mode persisted as a state in which the flag passively responded to bluff-body driven vortex shedding. Moreover, within large-amplitude flapping the same vortex-shedding and advection processes were found to persist even when the flapping amplitude varied by up to roughly $25\%$ from the uniform-stiffness flag. This outcome demonstrates that the bending profile under these different distributions is modulated to accommodate consistent FSI dynamics across a relatively wide range of stiffness distribution. This investigation opens questions about which bending profiles within large-amplitude flapping are optimal for energy capture, and how to incorporate a (potentially time varying) stiffness distribution to obtain desired FSI dynamics. This study considered a specific mass ratio and did not include an investigation into the effect of initial inclination angle of the flag. A characterization of these effects as well as the role of three dimensionality are avenues for future study.  

\section{Declaration of interest}

The author reports no conflict of interest.

\bibliographystyle{jfm}
\bibliography{inv_bib}{}

\begin{thebibliography}{17}
\expandafter\ifx\csname natexlab\endcsname\relax\def\natexlab#1{#1}\fi
\def\au#1{#1} \def\ed#1{#1} \def\yr#1{#1}\def\at#1{#1}\def\jt#1{\textit{#1}}
  \def\bt#1{#1}\def\bvol#1{\textbf{#1}} \def\vol#1{#1} \def\pg#1{#1}
  \def\publ#1{#1}\def\arxiv#1{#1}\def\org#1{#1}\def\st#1{\textit{#1}}

\bibitem[Floryan \& Rowley(2020)]{floryan2020distributed}
{\sc \au{Floryan, Daniel} \& \au{Rowley, Clarence~W}} \yr{2020}
  \at{Distributed flexibility in inertial swimmers}.  \jt{Journal of Fluid
  Mechanics}  \bvol{888}.

\bibitem[Goza \& Colonius(2017)]{Goza2017}
{\sc \au{Goza, Andres} \& \au{Colonius, Tim}} \yr{2017}  \at{A strongly-coupled
  immersed-boundary formulation for thin elastic structures}.  \jt{Journal of
  Computational Physics}  \bvol{336},  \pg{401--411}.

\bibitem[Goza {\em et~al.\/}(2018)Goza, Colonius \& Sader]{Goza2018}
{\sc \au{Goza, Andres}, \au{Colonius, Tim} \& \au{Sader, John~E}} \yr{2018}
  \at{Global modes and nonlinear analysis of inverted-flag flapping}.
  \jt{Journal of Fluid Mechanics}  \bvol{857},  \pg{312--344}.

\bibitem[Gurugubelli \& Jaiman(2015)]{Gurugubelli2015}
{\sc \au{Gurugubelli, PS} \& \au{Jaiman, RK}} \yr{2015}  \at{Self-induced
  flapping dynamics of a flexible inverted foil in a uniform flow}.
  \jt{Journal of Fluid Mechanics}  \bvol{781},  \pg{657--694}.

\bibitem[Gurugubelli \& Jaiman(2019)]{gurugubelli2019large}
{\sc \au{Gurugubelli, Pardha~S} \& \au{Jaiman, Rajeev~K}} \yr{2019}  \at{Large
  amplitude flapping of an inverted elastic foil in uniform flow with spanwise
  periodicity}.  \jt{Journal of Fluids and Structures}  \bvol{90},
  \pg{139--163}.

\bibitem[Huertas-Cerdeira {\em et~al.\/}(2021)Huertas-Cerdeira, Goza, Sader,
  Colonius \& Gharib]{huertas2021dynamics}
{\sc \au{Huertas-Cerdeira, Cecilia}, \au{Goza, Andres}, \au{Sader, John~E},
  \au{Colonius, Tim} \& \au{Gharib, Morteza}} \yr{2021}  \at{Dynamics of an
  inverted cantilever plate at moderate angle of attack}.  \jt{Journal of Fluid
  Mechanics}  \bvol{909}.

\bibitem[Kim {\em et~al.\/}(2013)Kim, Coss{\'e}, Cerdeira \& Gharib]{Kim2013}
{\sc \au{Kim, Daegyoum}, \au{Coss{\'e}, Julia}, \au{Cerdeira, Cecilia~Huertas}
  \& \au{Gharib, Morteza}} \yr{2013}  \at{Flapping dynamics of an inverted
  flag}.  \jt{Journal of Fluid Mechanics}  \bvol{736}.

\bibitem[Orrego {\em et~al.\/}(2017)Orrego, Shoele, Ruas, Doran, Caggiano,
  Mittal \& Kang]{orrego2017harvesting}
{\sc \au{Orrego, Santiago}, \au{Shoele, Kourosh}, \au{Ruas, Andre}, \au{Doran,
  Kyle}, \au{Caggiano, Brett}, \au{Mittal, Rajat} \& \au{Kang, Sung~Hoon}}
  \yr{2017}  \at{Harvesting ambient wind energy with an inverted piezoelectric
  flag}.  \jt{Applied energy}  \bvol{194},  \pg{212--222}.

\bibitem[Ryu {\em et~al.\/}(2015)Ryu, Park, Kim \& Sung]{Ryu2015}
{\sc \au{Ryu, Jaeha}, \au{Park, Sung~Goon}, \au{Kim, Boyoung} \& \au{Sung,
  Hyung~Jin}} \yr{2015}  \at{Flapping dynamics of an inverted flag in a uniform
  flow}.  \jt{Journal of Fluids and Structures}  \bvol{57}.

\bibitem[Sader {\em et~al.\/}(2016{\natexlab{{\em a\/}}})Sader, Coss{\'e}, Kim,
  Fan \& Gharib]{Sader2016a}
{\sc \au{Sader, John~E}, \au{Coss{\'e}, Julia}, \au{Kim, Daegyoum}, \au{Fan,
  Boyu} \& \au{Gharib, Morteza}} \yr{2016{\natexlab{{\em a\/}}}}
  \at{Large-amplitude flapping of an inverted flag in a uniform steady flow--a
  vortex-induced vibration}.  \jt{Journal of Fluid Mechanics}  \bvol{793}.

\bibitem[Sader {\em et~al.\/}(2016{\natexlab{{\em b\/}}})Sader,
  Huertas-Cerdeira \& Gharib]{Sader2016b}
{\sc \au{Sader, John~E}, \au{Huertas-Cerdeira, Cecilia} \& \au{Gharib,
  Morteza}} \yr{2016{\natexlab{{\em b\/}}}}  \at{Stability of slender inverted
  flags and rods in uniform steady flow}.  \jt{Journal of Fluid Mechanics}
  \bvol{809},  \pg{873--894}.

\bibitem[Sarpkaya(2004)]{Sarpkaya2004}
{\sc \au{Sarpkaya, Turgut}} \yr{2004}  \at{A critical review of the intrinsic
  nature of vortex-induced vibrations}.  \jt{Journal of Fluids and Structures}
  \bvol{19},  \pg{389--447}.

\bibitem[Shoele \& Mittal(2016)]{Shoele2016}
{\sc \au{Shoele, Kourosh} \& \au{Mittal, Rajat}} \yr{2016}  \at{Energy
  harvesting by flow-induced flutter in a simple model of an inverted
  piezoelectric flag}.  \jt{Journal of Fluid Mechanics}  \bvol{790},
  \pg{582--606}.

\bibitem[Tavallaeinejad {\em et~al.\/}(2020{\natexlab{{\em
  a\/}}})Tavallaeinejad, Pa{\"\i}doussis, Legrand \&
  Kheiri]{tavallaeinejad2020instability}
{\sc \au{Tavallaeinejad, Mohammad}, \au{Pa{\"\i}doussis, Michael}, \au{Legrand,
  Mathias} \& \au{Kheiri, Mojtaba}} \yr{2020{\natexlab{{\em a\/}}}}
  \at{Instability and the post-critical behaviour of two-dimensional inverted
  flags in axial flow}.  \jt{Journal of Fluid Mechanics}  \bvol{890}.

\bibitem[Tavallaeinejad {\em et~al.\/}(2020{\natexlab{{\em
  b\/}}})Tavallaeinejad, Pa{\"\i}doussis, Salinas, Legrand, Kheiri \&
  Botez]{tavallaeinejad2020flapping}
{\sc \au{Tavallaeinejad, Mohammad}, \au{Pa{\"\i}doussis, Michael}, \au{Salinas,
  Manuel~Flores}, \au{Legrand, Mathias}, \au{Kheiri, Mojtaba} \& \au{Botez,
  Ruxandra}} \yr{2020{\natexlab{{\em b\/}}}}  \at{Flapping of heavy inverted
  flags: a fluid-elastic instability}.  \jt{Journal of Fluid Mechanics}
  \bvol{904}.

\bibitem[Tavallaeinejad {\em et~al.\/}(2021)Tavallaeinejad, Salinas,
  Pa{\"\i}doussis, Legrand, Kheiri \& Botez]{tavallaeinejad2021dynamics}
{\sc \au{Tavallaeinejad, Mohammad}, \au{Salinas, Manuel~Flores},
  \au{Pa{\"\i}doussis, Michael~P}, \au{Legrand, Mathias}, \au{Kheiri, Mojtaba}
  \& \au{Botez, Ruxandra~M}} \yr{2021}  \at{Dynamics of inverted flags:
  Experiments and comparison with theory}.  \jt{Journal of Fluids and
  Structures}  \bvol{101},  \pg{103199}.

\bibitem[Williamson \& Govardhan(2004)]{Williamson2004}
{\sc \au{Williamson, Charles~HK} \& \au{Govardhan, R.}} \yr{2004}
  \at{Vortex-induced vibrations}.  \jt{Annual Review of Fluid Mechanics}
  \bvol{36},  \pg{413--455}.

\end{thebibliography}

\end{document}